\documentclass[letterpaper,
twocolumn,
showpacs,
prb,
superscriptaddress,email
,showkeys%
]{revtex4-1}

\usepackage[pdftex]{graphicx}
\usepackage{graphicx}
\usepackage{mathrsfs}
\usepackage{amssymb}
\usepackage{amsmath,amsfonts,latexsym}
\usepackage{array,tabularx,color}
\usepackage{dcolumn}        
\usepackage[normalem]{ulem}
\usepackage{comment}
\usepackage{bm}

\begin{document}


\title{Ignition and formation dynamics of a polariton condensate on a semiconductor microcavity pillar}

\author{C. Ant\'{o}n}
\affiliation{Departamento de F\'isica de Materiales, Universidad Aut\'onoma de Madrid, Madrid 28049, Spain}
\affiliation{Instituto de Ciencia de Materiales ``Nicol\'as Cabrera'', Universidad Aut\'onoma de Madrid, Madrid 28049, Spain}

\author{D. Solnyshkov}
\affiliation{Institut Pascal, PHOTON-N2, Clermont Universit\'e, Blaise Pascal University, CNRS,24 Avenue des Landais, 63177 Aubi\`ere Cedex, France}

\author{G. Tosi}
\affiliation{Departamento de F\'isica de Materiales, Universidad Aut\'onoma de Madrid, Madrid 28049, Spain}

\author{M. D. Mart\'{i}n}
\affiliation{Departamento de F\'isica de Materiales, Universidad Aut\'onoma de Madrid, Madrid 28049, Spain}
\affiliation{Instituto de Ciencia de Materiales ``Nicol\'as Cabrera'', Universidad Aut\'onoma de Madrid, Madrid 28049, Spain}

\author{Z. Hatzopoulos}
\affiliation{FORTH-IESL, P.O. Box 1385, 71110 Heraklion, Crete, Greece}
\affiliation{Department of Physics, University of Crete, 71003 Heraklion, Crete, Greece}

\author{G. Deligeorgis}
\affiliation{FORTH-IESL, P.O. Box 1385, 71110 Heraklion, Crete, Greece}

\author{P.G. Savvidis}
\affiliation{FORTH-IESL, P.O. Box 1385, 71110 Heraklion, Crete, Greece}
\affiliation{Department of Materials Science and Technology, Univ. of Crete, 71003 Heraklion, Crete, Greece}

\author{G. Malpuech}
\affiliation{Institut Pascal, PHOTON-N2, Clermont Universit\'e, Blaise Pascal University, CNRS,24 Avenue des Landais, 63177 Aubi\`ere Cedex, France}

\author{L. Vi\~{n}a}
\email{luis.vina@uam.es}
\affiliation{Departamento de F\'isica de Materiales, Universidad Aut\'onoma de Madrid, Madrid 28049, Spain}
\affiliation{Instituto de Ciencia de Materiales ``Nicol\'as Cabrera'', Universidad Aut\'onoma de Madrid, Madrid 28049, Spain}
\affiliation{Instituto de F\'isica de la Materia Condensada, Universidad Aut\'onoma de Madrid, Madrid 28049, Spain}

\date{\today}

\begin{abstract}

We present an experimental study on the ignition and decay of a polariton optical parametric oscillator (OPO) in a semiconductor microcavity pillar. The combination of a continuous wave laser \emph{pump}, under quasi-phase matching conditions, and a non-resonant, 2 ps-long pulse \emph{probe} allows us to obtain the full dynamics of the system. The arrival of the \emph{probe} induces a blue-shift in the polariton emission, bringing the OPO process into resonance with the \emph{pump}, which triggers the OPO-process. We time-resolve the polariton OPO signal emission for more than 1 nanosecond in both real and momentum-space. We fully characterize the emission of the OPO signal with spectral tomography techniques. Our interpretations are backed up by theoretical simulations based on the 2D coupled Gross-Pitaevskii equation for excitons and photons.

\end{abstract}



\maketitle


\section{Introduction}
\label{sec:intro}

Exciton-polaritons in semiconductor microcavities (MCs), when injected with a \emph{pump} laser close to the inflection point of the lower polariton branch (LPB) dispersion, undergo a nonlinear process above a pump power threshold. Carrier-carrier interactions self-stimulate a coherent scattering from the \emph{pump} state into \emph{signal} and \emph{idler} polariton states, whose frequency and in-plane momentum fulfill phase-matching conditions. The \emph{signal} state population, generated by this optical parametric oscillator (OPO)~\cite{Savvidis2000,savvidis00:prb,Baumberg:2000qf,Saba:2001aa} process, reaches occupation values above one, exhibiting a new form of non-equilibrium superfluid behavior~\cite{amo2009}, metastability of quantum vortices~\cite{Krizhanovskii2010,Guda:2013aa} and persistence of currents~\cite{Sanvitto2010,Tosi2011a,Anton:12}.

Lateral etching of planar MCs has been successfully exploited for the creation of a new, large variety of geometries: the resulting discretization of the energy spectrum opens an interesting scenario of different OPO phase-matching conditions in one- (1D) and zero-dimensional (0D) MCs. In the former case, the 1D discretization of the LPB in several energy sub-bands yields the opportunity to obtain exotic intra- and inter-branch OPO processes.~\cite{Dasbach:2002tg,Abbarchi:2011ai,Lecomte:2013aa,PhysRevB.88.235312} Furthermore, interesting studies of the second order coherence of both signal and idler states have been performed recently.~\cite{Ardizzone:2012aa} In the latter case (0D OPO polaritons), although the most common excitation scheme for micro pillars is non-resonant excitation,~\cite{El-Daif:2006rr,bajoni2008,Maragkou:2010qe,Nardin:2009nx,Nardin:2010wd} different groups have reported the possibility to induce a parametric oscillation between discrete energy states: from the initial injected energy mode to two neighboring, signal and idler states.~\cite{Dasbach:2001lq,bajoni:051107,ferrier:031105}

In this work, we study a 40 $\mu$m-$\varnothing$ pillar MC, which is sufficiently large to neglect 0D confinement effects, but still with a bounded spatial extension where polaritons cannot propagate large distances. We excite the sample using a new \emph{pump+probe} excitation scheme, that differs from common resonant excitation~\cite{Savvidis2000,savvidis00:prb,Baumberg:2000qf,Saba:2001aa,Ballarini:2009aa}. We ignite a long lived OPO polariton signal and observe a transient behavior, characterized by a collective oscillation of the condensed OPO polaritons in the pillar. Thereafter, they reach a quasi-steady state displaying a ring-like emission pattern, due to repulsive interaction with the exciton population in the center of the pillar.~\cite{Kalevich14} The OPO signal is switched on with the pulsed \emph{probe} at the exciton energy level which blueshifts the LPB making the continuous wave (\emph{cw}) \emph{pump} enter in resonance conditions in a OPO-process that lives for $\sim$1 ns. We study the full dynamics of the creation and decay of this confined OPO condensate in real and momentum-space ($\mathbf{k}$-space). The interpretation of the experimental measurements of the spatial emission dynamics is supported by theoretical simulations using the 2D coupled Gross-Pitaevskii equations for excitons and photons.


\section{Sample and experimental setup}
\label{sec:sample}

\begin{figure}[htb]
\setlength{\abovecaptionskip}{-5pt}
\setlength{\belowcaptionskip}{-2pt}
\begin{center}
\includegraphics[width=.9\linewidth,angle=0]{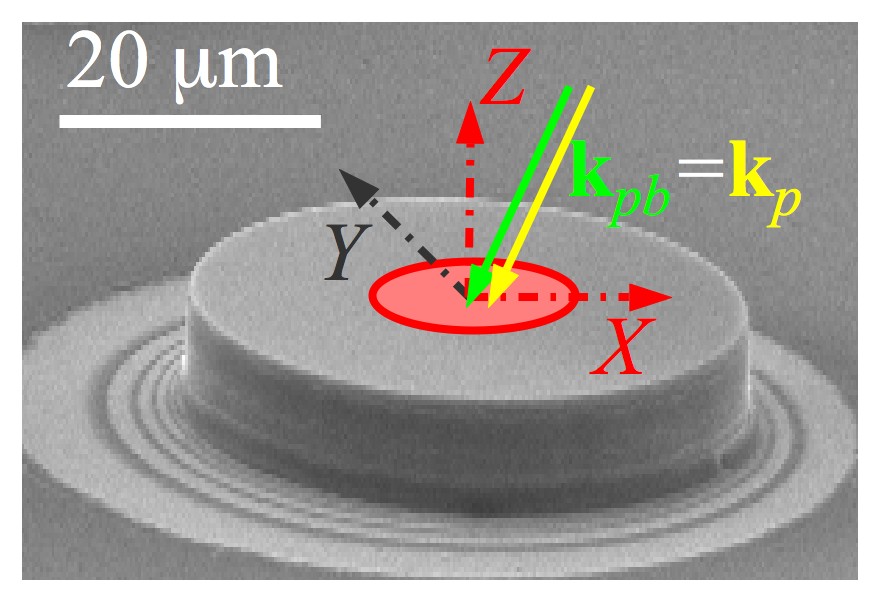}
\end{center}
\caption {(Color online) Scanning electron microscopy image of a 40 $\mu$m-$\varnothing$ pillar with the \emph{pump}+\emph{probe} excitation scheme: the full arrows, contained in the $XZ$ plane, sketch the beams directions, $\mathbf{k}_{pb}$ and $\mathbf{k}_{p}$, impinging at the center of the pillar. Three dot-dashed arrows define the coordinates origin chosen for the experiments.}
\label{fig:pillar_scheme}
\end{figure}

We investigate a high-quality $5\lambda/2$ AlGaAs-based MC with 12 embedded quantum wells, and a Rabi splitting $\Omega_R = 9~$ meV. Further information about this sample is given in Ref.~\onlinecite{Tsotsis:2012qy}. Pillars with different diameters have been sculpted through reactive ion etching. We have chosen a 40 $\mu$m-$\varnothing$ pillar in an area of the sample where the detuning is close to zero.\cite{detunning} Figure~\ref{fig:pillar_scheme} shows a scanning electron microscopy image of such a pillar, including the excitation scheme.

\begin{figure}[htb]
\setlength{\abovecaptionskip}{-5pt}
\setlength{\belowcaptionskip}{-2pt}
\begin{center}
\includegraphics[trim=0.4cm 0.3cm 0.4cm 0.2cm, clip=true,width=1\linewidth,angle=0]{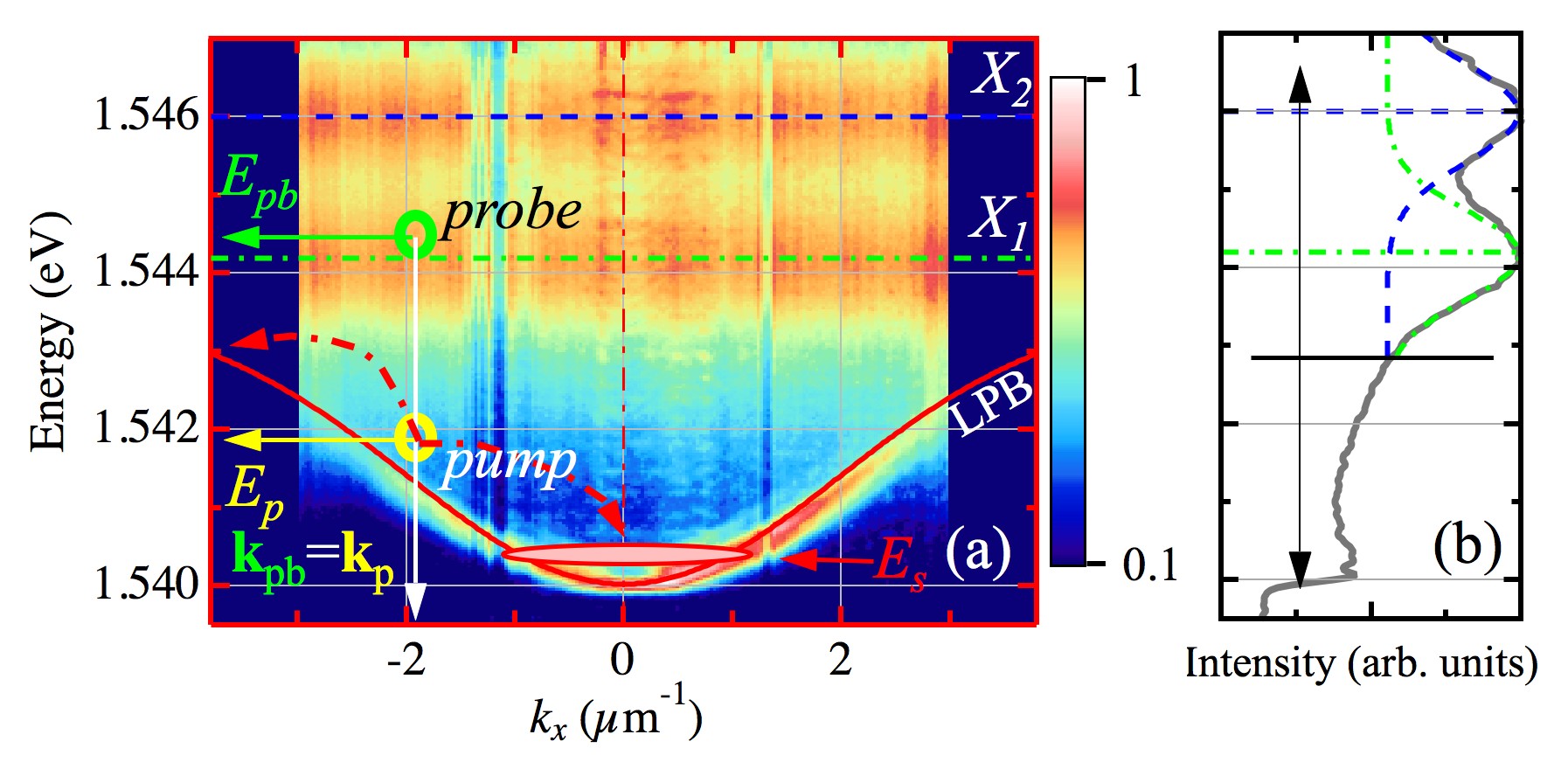}
\end{center}
\caption {(Color online) (a) Experimental LPB dispersion obtained under non-resonant (1.612 eV), weak excitation conditions; the overlapping thin, full curve is a guide to the eye. The horizontal dot-dashed and dashed lines mark the energies 1.544 eV and 1.546 eV of the excitons $\mathbf{X_1}$ and $\mathbf{X_2}$, respectively. The energy, $E_p/E_{pb}$, and momentum, $\mathbf{k}_{p}$/$\mathbf{k}_{pb}$, of the \emph{pump}/\emph{probe} is marked with a circle and left-pointing arrows. Dot-dashed arrows depict the OPO process that yields the OPO signal at $E_s$ and $\mathbf{k}_{s} \approx 0$, marked with an elongated circle. The PL is coded in a false, normalized, logarithmic color scale. (b) Spectrum emission integrated in $\mathbf{k}$-space depicted in a full line. The two bare exciton levels are obtained with a double gaussian fit: the dot-dashed (dashed) horizontal line marks the energy of the maximum PL emitted by the $X_1$ ($X_2$) exciton. A horizontal full line schematically separates, at 1.543 eV, the spectrum that correspond to excitons (above, upward pointing arrow) and polaritons (below, downward pointing arrow) in the pillar.}
\label{fig:disrel}
\end{figure}

The sample, mounted in a cold-finger cryostat and kept at 10 K, is excited with \emph{pump} and \emph{probe} laser beams under the following conditions. For the continuous wave experiments, Section~\ref{subsec:cwopo}, we use only a \emph{pump} beam obtained from a \emph{cw} Ti:Al$_2$O$_3$ laser. It is tuned at $E_p=1.5416$~eV, impinging on the sample with an in-plane momentum  $(\mathbf{k}_p)_x=-1.9$~$\mu$m$^{-1}$, fulfilling the phase-matching conditions $2E_p = E_s+E_i$ and $2\mathbf{k}_p=\mathbf{k}_s+\mathbf{k}_i$ (where the subindex $s$/$i$ means \emph{signal}/\emph{idler}). Its power is set to $P_p=$ 160 mW. For the time-resolved experiments, Subsecs.~\ref{subsec:p+pb} and \ref{subsec:pb}, the excitation scheme is represented in Fig.~\ref{fig:disrel}. In this case the \emph{cw pump} beam is out of OPO phase-matching conditions, since now its energy is tuned slightly above the LPB ($E_p=1.5419$~eV). The second excitation source, \emph{probe}, is a pulsed Ti:Al$_2$O$_3$ laser (2 ps-long pulses); it is tuned into resonance with the exciton mode ($E_{pb}=1.5445$~eV), and its power, $P_{pb}= $ 230 mW, is strong enough to trigger the OPO-process together with the \emph{pump}. The origin of time $t=0$ is set to the instant when the \emph{probe} impinges on the pillar. The two excitonic lines labelled $X_1$ and $X_2$ originate from excitons uncoupled to the cavity modes, due to slight quantum well thickness variations, of the order of a monolayer, between different quantum wells. Further information about the origin of the exciton emission is given in Ref. \onlinecite{Kalevich14}.

For the experimental results described in Sec.~\ref{sec:experiments}, the laser beams are focused on the sample through a high numerical-aperture (0.6) lens, forming two overlapping elliptically shaped spots (10/20 $\mu$m-$\varnothing$ minor/major axis along the Y/X axis) impinging on the pillar with an in-plane momentum $\mathbf{k}_{p/pb}=\left\{(\mathbf{k}_{p/pb})_x,(\mathbf{k}_{p/pb})_y\right\}=\left\{-1.9, 0\right\}$~$\mu$m$^{-1}$, see full arrows in Fig.~\ref{fig:pillar_scheme}. The same lens is used to collect and direct the emission towards a 0.5 m spectrometer coupled to a CCD (Section~\ref{subsec:cwopo}) and a streak camera (Subsecs.~\ref{subsec:p+pb} and \ref{subsec:pb}). The photoluminescence (PL) can be resolved in the near- (real-space) as well as in the far-field ($\mathbf{k}$-space). The distribution of polaritons in $\mathbf{k}$-space is  accessed by imaging the Fourier plane of the lens used to collect the PL, taking advantage of the direct relation between the angle of emission and the in-plane momentum of polaritons.\cite{richard05} To avoid the direct reflection of the \emph{pump} and \emph{probe} beams, we block the emission in $\mathbf{k}$-space for $\left|\mathbf{k}\right| > 1.5$~$\mu$m$^{-1}$ and we spectrally filter the polariton PL with an energy detection range of 1 meV around 1.54 eV. The lens that focuses the real or $\mathbf{k}$-space PL distribution into the spectrometer entrance slit is displaced laterally by discrete steps, yielding a tomographic reconstruction of energy- and time-resolved images.

\section{Experimental results and discussion}
\label{sec:experiments}

This section compiles the results of \emph{cw} and time-resolved experiments in the pillar, which are organized as follows. In Section~\ref{subsec:cwopo} we address the tomographic spectral distribution of polaritons in the pillar under \emph{cw}-OPO excitation, in both real- and $\mathbf{k}$-space. In Section~\ref{subsec:p+pb} we study the dynamics of the polariton emission under \emph{pump}$+$\emph{probe} excitation. In this case, the \emph{cw} \emph{pump} laser beam is out of OPO-conditions (slightly blue-detuned from the LPB); only after the arrival of the \emph{probe}, the induced blue-shift of the LPB is large enough to start the OPO-process. In Section~\ref{subsec:pb}, for the sake of completeness, we present the polariton dynamics with the \emph{probe} excitation only, where the decay of the polariton PL is faster and limited by the relaxation dynamics from photo-generated excitons towards the polariton ground state.

\subsection{CW spectroscopy characterization of OPO signal emission}
\label{subsec:cwopo}

\begin{figure}[!htb]
\setlength{\abovecaptionskip}{-5pt}
\setlength{\belowcaptionskip}{-2pt}
\begin{center}
\includegraphics[trim=0.4cm 0.3cm 0.3cm 0.2cm, clip=true,width=1.0\linewidth,angle=0]{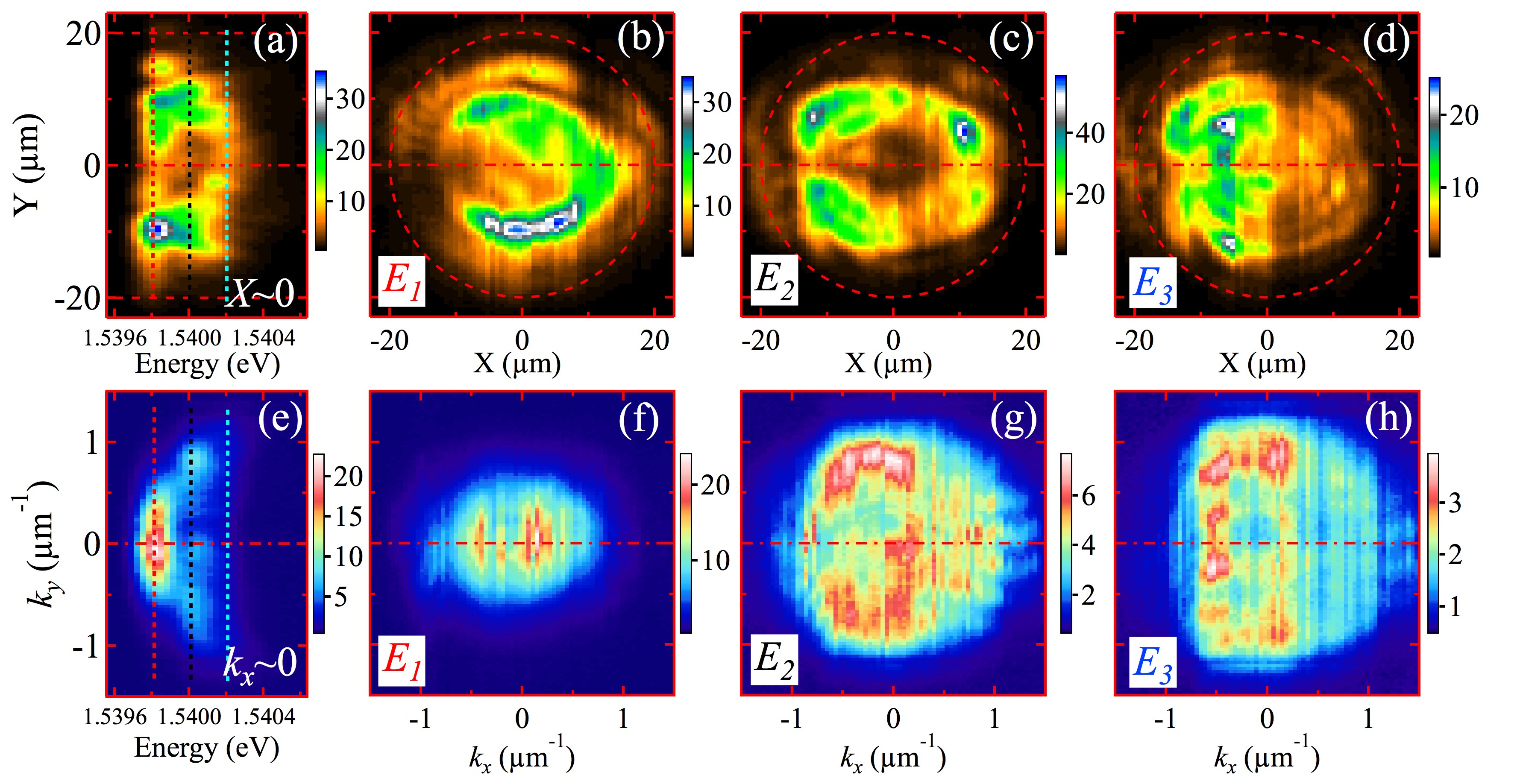}
\vspace{-0.5cm}
\end{center}
\caption{(Color online) (a)/(e) Real-/Momentum-space polariton spectrum emission at $X=0$/$k_x=0$ cross-section in the pillar. The three, dashed, vertical lines mark the position where the full $X-Y$/$k_x-k_y$ PL map has been reconstructed in panels (b-d)/(f-h). The three selected energies are: $E_{1}= 1.5398$~eV, $E_{2}= 1.5400$~eV and $E_{3}= 1.5402$~eV. In panels (b-d) dashed, red circles mark the limits of the pillar. A vertical red line marks the resolved cross-section of the spectrum shown in panels (a)/(d) for real-/$\mathbf{k}$-space. The PL maps are coded in linear, false color scales.}
\label{fig:CWmodes}
\end{figure}

Figure~\ref{fig:CWmodes} summarizes the spectral tomography of the polariton emission in both real- and $\mathbf{k}$-space under a \emph{cw} \emph{pump} excitation. The OPO signal energy is $E_s\approx 1.54$~eV, with a full width $\Delta E_s \approx 0.8$~meV.
Figure~\ref{fig:CWmodes}(a)/(e) shows the polariton emission in real-/$\mathbf{k}$-space ($Y$/$k_y$) at the central $X=0$/$k_x=0$ cross-section. Since the \emph{pump} is impinging at the center of the pillar, the polariton emission displays a ring-like distribution, as can be seen in the cross section in Fig.~\ref{fig:CWmodes}(a), where the emission is spatially enclosed in the area $5<\left|\mathbf{r}\right|<15$~$\mu$m. Figure~\ref{fig:CWmodes}(e) shows that polaritons are static at low energies ($\sim$1.5398 eV), since their emission is confined at $\mathbf{k}\sim 0$, and they spread in a disk of radius $|\mathbf{k}|<1.2$~$\mu$m$^{-1}$ at slightly higher energies ($\sim$1.540 eV).

We perform a two-dimensional reconstruction of the spectrum emission, showing the full $X-Y$/$k_x-k_y$ PL distribution of polaritons at three, selected energies: $E_{1}= 1.5398$~eV, $E_{2}= 1.5400$~eV and $E_{3}= 1.5402$~eV, Figs.~\ref{fig:CWmodes}(b-d)/(f-h). Figure~\ref{fig:CWmodes}(b) reveals the ring-like distribution of polaritons at $E_1$. The angle of incidence of the \emph{pump} creates an asymmetrical blueshift in the left side of the pillar due to polariton propagation, so the polariton ring-shaped emission is broken in the region $\left\{x,y\right\}=\left\{-10, 0\right\}$~$\mu$m. At a higher energy $E_2$, Fig.~\ref{fig:CWmodes}(c), polaritons emit from a ring of radius 10~$\mu$m with a smaller side gap. Figure~\ref{fig:CWmodes}(d) shows an almost flat disk of emission, whose radius is $\sim$15~$\mu$m.

In $\mathbf{k}$-space, Fig.~\ref{fig:CWmodes}(f) shows a small disk (radius 0.7~$\mu$m$^{-1}$) in the polariton distribution. This demonstrates that polaritons lying at low energy ($E_{1}$), which are distributed along the ring in Fig. ~\ref{fig:CWmodes}(b), are confined in real space, with a small amount of kinetic energy. The confinement of polaritons at $E_{1}$ is also hinted in the momentum cross-section in Fig. ~\ref{fig:CWmodes}(e). Figures~\ref{fig:CWmodes}(g) and \ref{fig:CWmodes}(h) evidence a flat distribution in momentum space, where polaritons have all possible values of momenta inside a disk of radius $\left|\mathbf{k}\right|<1$ and $1.5$~$\mu$m$^{-1}$, respectively. For energies higher than $E_3$, polariton emission is distributed in a ring (not shown) corresponding to a cloud of uncondensed, hot polaritons in the LPB.

\subsection{Igniting a long-living OPO-process with a \emph{probe}-induced blueshift}
\label{subsec:p+pb}

In this section we show how the OPO process is triggered by the arrival of a \emph{probe} beam. Moreover, through real-space measurements, we reveal the ring-shape distribution of \emph{signal} polaritons and, analyzing $\mathbf{k}$-space images, we characterize their movement around the pillar.

\begin{figure*}[htb]
\setlength{\abovecaptionskip}{-5pt}
\setlength{\belowcaptionskip}{-2pt}
\begin{center}
\includegraphics[trim=0.5cm 0.8cm 0.3cm 0.2cm, clip=true,width=1\linewidth,angle=0]{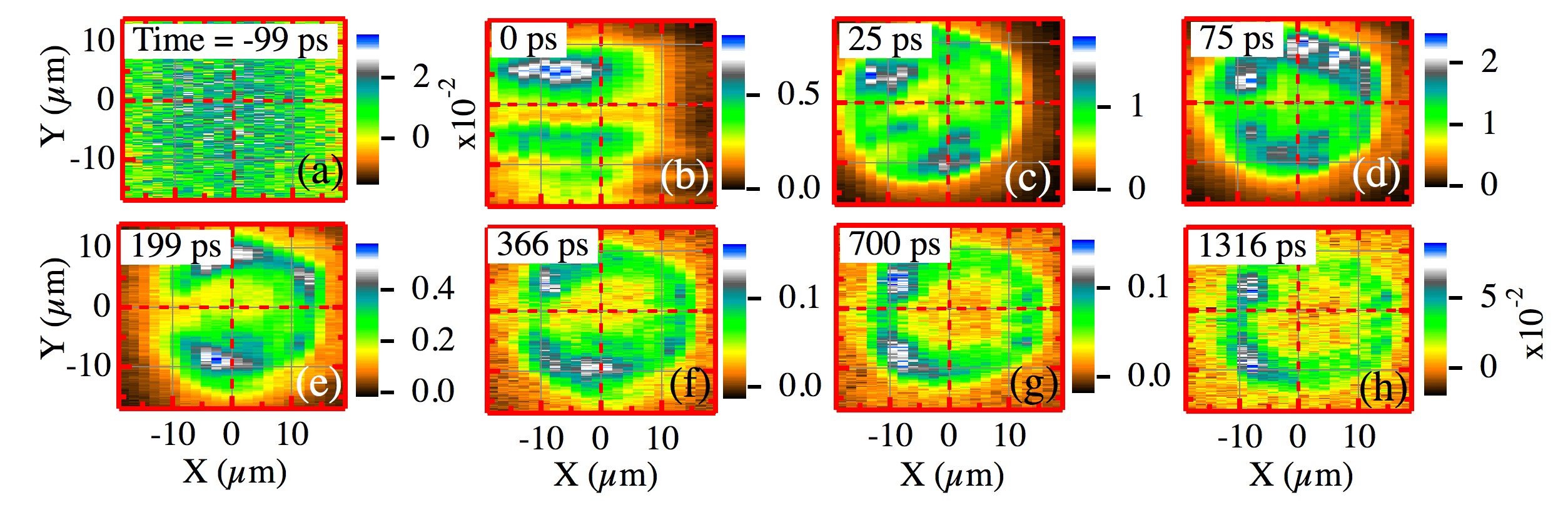}
\end{center}
\caption{(Color online) Real-space polariton emission dynamics in the pillar under \emph{pump} and \emph{probe} beam excitation. The time is shown by the labels in each panel. The time $t=0$ coincides with the arrival of the \emph{probe}. The PL is coded in a linear, false color scale shown on the right of each panel. The complete $X-Y$ polariton emission dynamics is available as Supplemental Material.~\cite{supplemat}}
\label{fig:realCWpulsed}
\end{figure*}

\begin{figure*}[htb]
\setlength{\abovecaptionskip}{-5pt}
\setlength{\belowcaptionskip}{-2pt}
\begin{center}
\includegraphics[trim=0.5cm 0.8cm 0.3cm 0.2cm, clip=true,width=1\linewidth,angle=0]{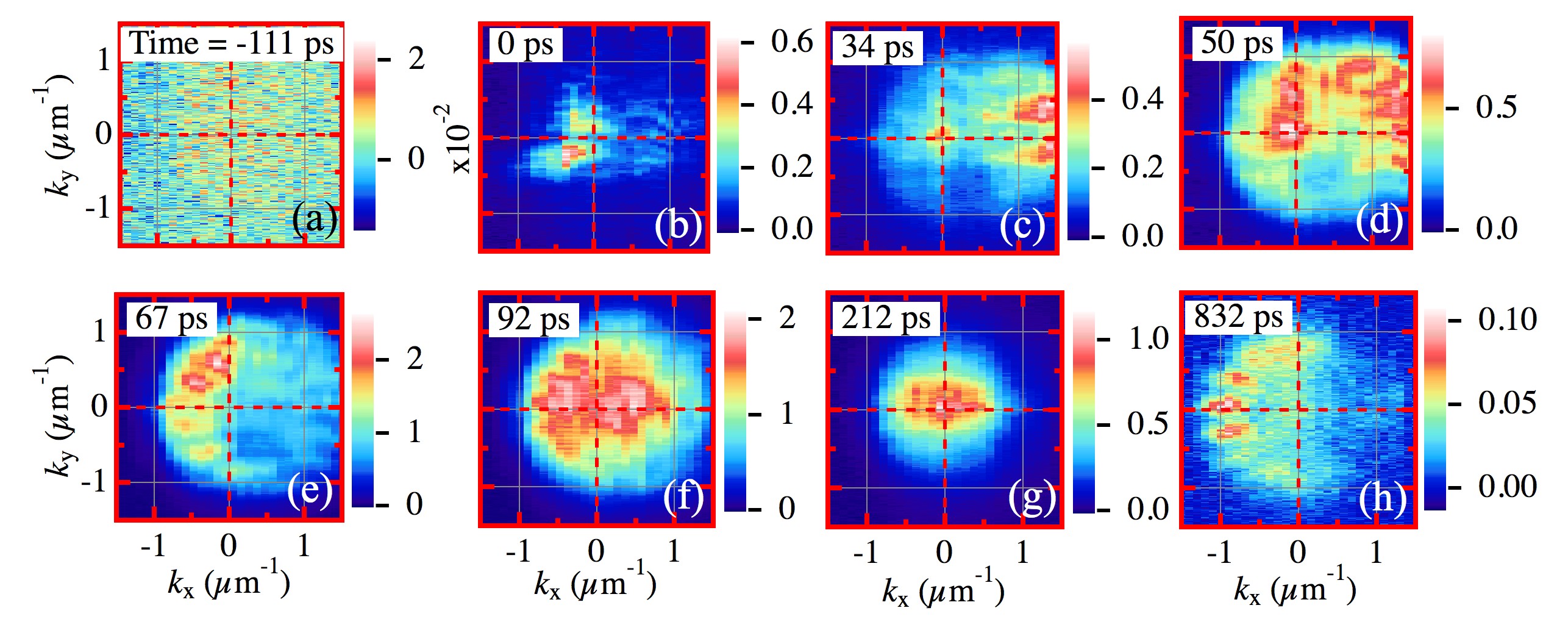}
\end{center}
\caption{(Color online) Momentum-space polariton emission dynamics in the pillar. Same excitation conditions as in Fig.~\ref{fig:realCWpulsed}. The time is shown by the labels in each panel. The PL is coded in a linear, false color scale shown on the right of each panel. The complete $k_X-k_Y$ polariton emission dynamics is available as Supplemental Material.~\cite{supplemat}}
\label{fig:momentumCWpulsed}
\end{figure*}

The \emph{pump} (\emph{cw}) $+$ \emph{probe} (pulsed) excitation configuration activates a long-lived stimulated OPO scattering process. The time during which the OPO is active is much longer than the photon lifetime, estimated from the Q-factor to be $\sim$ 10 ps. The real- and $\mathbf{k}$-space dynamics or the signal emission are presented in Figs.~\ref{fig:realCWpulsed} and~\ref{fig:momentumCWpulsed}, respectively. In these time-resolved experiments, the energy resolution is similar to that of the energy width of the OPO signal $\Delta E_s$.

Figure~\ref{fig:realCWpulsed} compiles two dimensional images of the polariton signal emission at different times in real-space. For each panel, the time is displayed at the left upper corner, being the temporal origin set at the instant when the \emph{probe} impinges on the sample. Fig.~\ref{fig:realCWpulsed}(a) shows that, before the arrival of the \emph{probe} pulse ($t=-99$~ps), there is no signal emission since the \emph{pump} is off of phase-matching conditions. At $t=0$, Fig.~\ref{fig:realCWpulsed}(b), the \emph{probe} impinges on the pillar; the spot shape is distorted due to the fact that $\mathbf{k}_{pb}\neq0$. The polariton emission rapidly arises from the whole pillar surface, seen as a flat homogeneous disk, Fig.~\ref{fig:realCWpulsed}(c). The PL increases during $\sim$100 ps, as shown in Fig.~\ref{fig:realCWpulsed}(d); thereafter, the extra-population of polaritons induced by the \emph{probe} decreases. When the OPO process has switched-on, the polariton dispersion becomes ring-shaped, Fig.~\ref{fig:realCWpulsed}(e): the real-space polariton distribution resembles that shown in Fig.~\ref{fig:CWmodes}(c). Polaritons emit close to the border of the pillar, due to the blueshift induced by the \emph{cw} \emph{pump}. This polariton emission persists for more than $\sim$1 ns, see Figs.~\ref{fig:realCWpulsed}(f-h).

Figure~\ref{fig:momentumCWpulsed} shows the dynamics in $\mathbf{k}$-space of the polariton population. As observed in real-space, there is no OPO signal emission before the arrival of the \emph{probe}, Fig.~\ref{fig:momentumCWpulsed}(a). The pulsed \emph{probe} arrives to the pillar at $t=0$, Fig.~\ref{fig:momentumCWpulsed}(b), the spurious emission from $\mathbf{k}\sim0$ is unfiltered scattered laser light. The \emph{probe} creates an extra polariton population around $\mathbf{k}=\left\{-1.5, 0\right\}$~$\mu$m$^{-1}$ that rapidly moves in the $+k_x$ direction (not shown here, see the movie in the Supplementary Material).~\cite{supplemat} At $t=34$~ps, Fig.~\ref{fig:momentumCWpulsed}(c), the population has moved towards $\mathbf{k}=\left\{1.5, 0\right\}$~$\mu$m$^{-1}$. Fig.~\ref{fig:momentumCWpulsed}(d) shows that, at $t=50$~ps, the polariton emission is homogeneously distributed in a disk of radius $\left|\mathbf{k}\right|\approx 1$~$\mu$m$^{-1}$. At $t=67$~ps, Fig.~\ref{fig:momentumCWpulsed}(e), the population reverts its angle of emission towards $-k_x$. As it was mentioned for Fig.~\ref{fig:realCWpulsed}, $\sim100$~ps after the arrival of the \emph{probe}, its induced extra-population weakens and the OPO signal is redistributed at the center of \textbf{k}-space, Fig.~\ref{fig:momentumCWpulsed}(f). The oscillations arise from the \emph{probe}-injected excitons, which are excited with a certain angle.The relaxation of excitons yields polaritons that possess a non-zero momentum. As clearly inferred from the dynamics of the $\mathbf{k}$-space distribution, polaritons change their initial momentum due to several bounces against the MC wall. The gain follows the injected \emph{probe} distribution as it moves within the pillar, until the excitonic population dies off and a more stable, switched-on OPO process takes place, which resembles the polariton distribution in real and $\mathbf{k}$-space shown in Fig. \ref{fig:CWmodes}. For longer times, $t=212$~ps (Fig.~\ref{fig:momentumCWpulsed}(g)), the emission is mainly perpendicular to the sample surface, i.e. at $\mathbf{k}\approx0$. At later times, Fig.~\ref{fig:momentumCWpulsed}(h), two effects are observed: firstly, there is a lobe-like structure at $\mathbf{k}=\left\{1.0, 0\right\}$~$\mu$m$^{-1}$ and, secondly, there is a progressive decay of the central ($\mathbf{k}=0$) population. The latter effect is due to the fact that the emission energy is red-shifting with time and we have not been following this red-shift with the streak camera since its energy detection is fixed for these experiments.


\subsection{Non-resonant excitation solely with a pulsed \emph{probe}}
\label{subsec:pb}

\begin{figure*}[!hbt]
\setlength{\abovecaptionskip}{-5pt}
\setlength{\belowcaptionskip}{-2pt}
\begin{center}
\includegraphics[trim=0.8cm 0.6cm 0.3cm 0.2cm, clip=true,width=1\linewidth,angle=0]{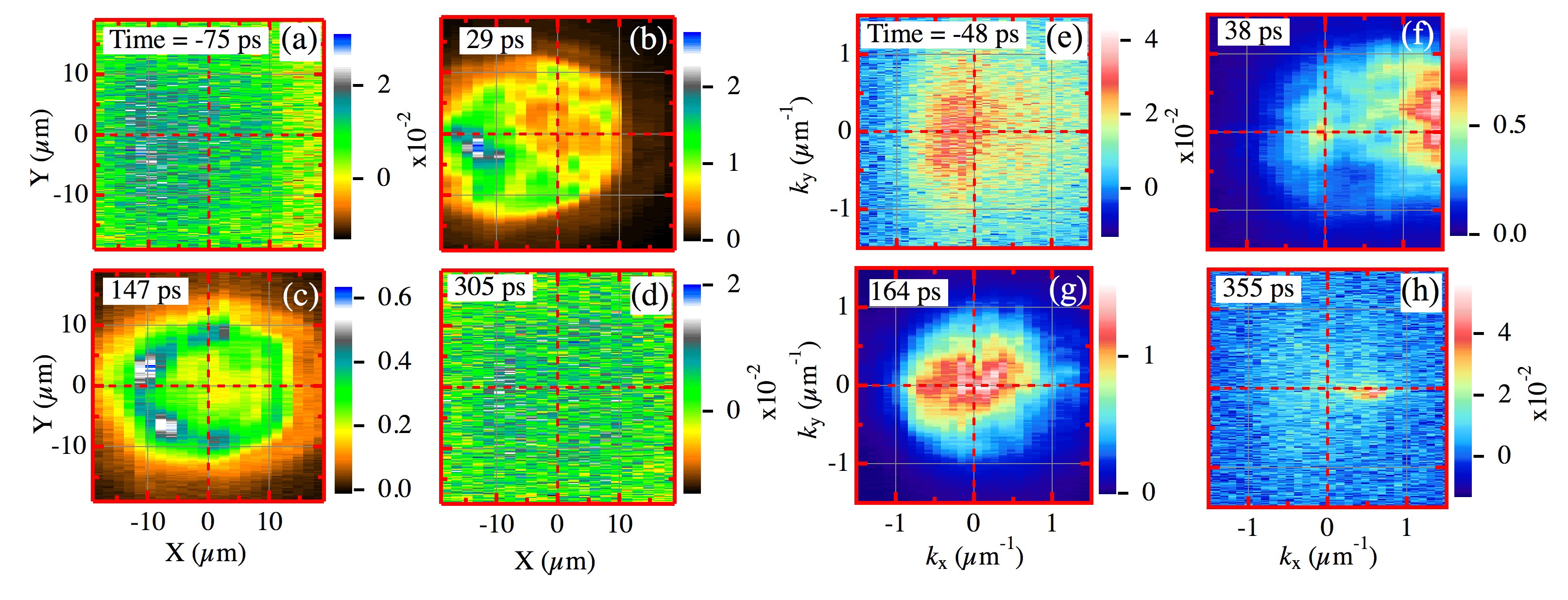}
\end{center}
\caption{(Color online) (a)-(d) Real/(e)-(h) $\mathbf{k}$-space polariton emission distribution in the pillar, the time is shown by the labels in each panel. Only the \emph{probe} beam excites at the center of the pillar. The time $t=0$ coincides with the arrival of the \emph{probe}. The PL is coded in a linear, false color scale shown on the right of each panel. The complete $X-Y$ and $k_X-k_Y$ polariton emission dynamics is available as Supplemental Material.~\cite{supplemat}}
\label{fig:onlypulsed}
\end{figure*}

In this section, for completeness, we address the polariton dynamics obtained when only the pulsed \emph{probe} beam excites the pillar. The polariton emission dynamics in real- and $\mathbf{k}$-space is summarized in Figs.~\ref{fig:onlypulsed}(a-d) and Figs.~\ref{fig:onlypulsed}(e-h), respectively. Fig.~\ref{fig:onlypulsed}(a) shows the absence of emission before the arrival of the \emph{probe}. Figs.~\ref{fig:onlypulsed}(b,c) show the switch-on of the polariton emission, following a similar dynamics to that shown in Fig.~\ref{fig:realCWpulsed}, during the first $\sim$100 ps. Fig.~\ref{fig:onlypulsed}(d) demonstrates the shorter lifetime of the polariton population created under these excitation conditions; after $\sim$250 ps, the polariton PL has disappeared.

In a similar fashion, Figs.~\ref{fig:onlypulsed}(e-h) show the dynamics of the emission in $\mathbf{k}$-space. An oscillation of the polariton momentum similar to that described in Fig.~\ref{fig:momentumCWpulsed}(c-e) is obtained here during the first $\sim$100 ps. The polariton emission moves in the $k_x$ direction, going from $-1.5$~$\mu$m$^{-1}$ to $+1.5$~$\mu$m$^{-1}$ in $\sim$70 ps. At later times, Fig.~\ref{fig:onlypulsed}(g), the emission originates from ${k} \approx 0$, and for $t=355$~ps, the lack of emission is confirmed, Fig.~\ref{fig:onlypulsed}(h).

\subsection{Comparison of the dynamics of the two excitation schemes involving a \emph{probe}}
\label{subsec:comp}

\begin{figure}[!htb]
\setlength{\abovecaptionskip}{-5pt}
\setlength{\belowcaptionskip}{-2pt}
\begin{center}
\includegraphics[trim=0.2cm 0.1cm 0.0cm 0.0cm, clip=true,width=1\linewidth,angle=0]{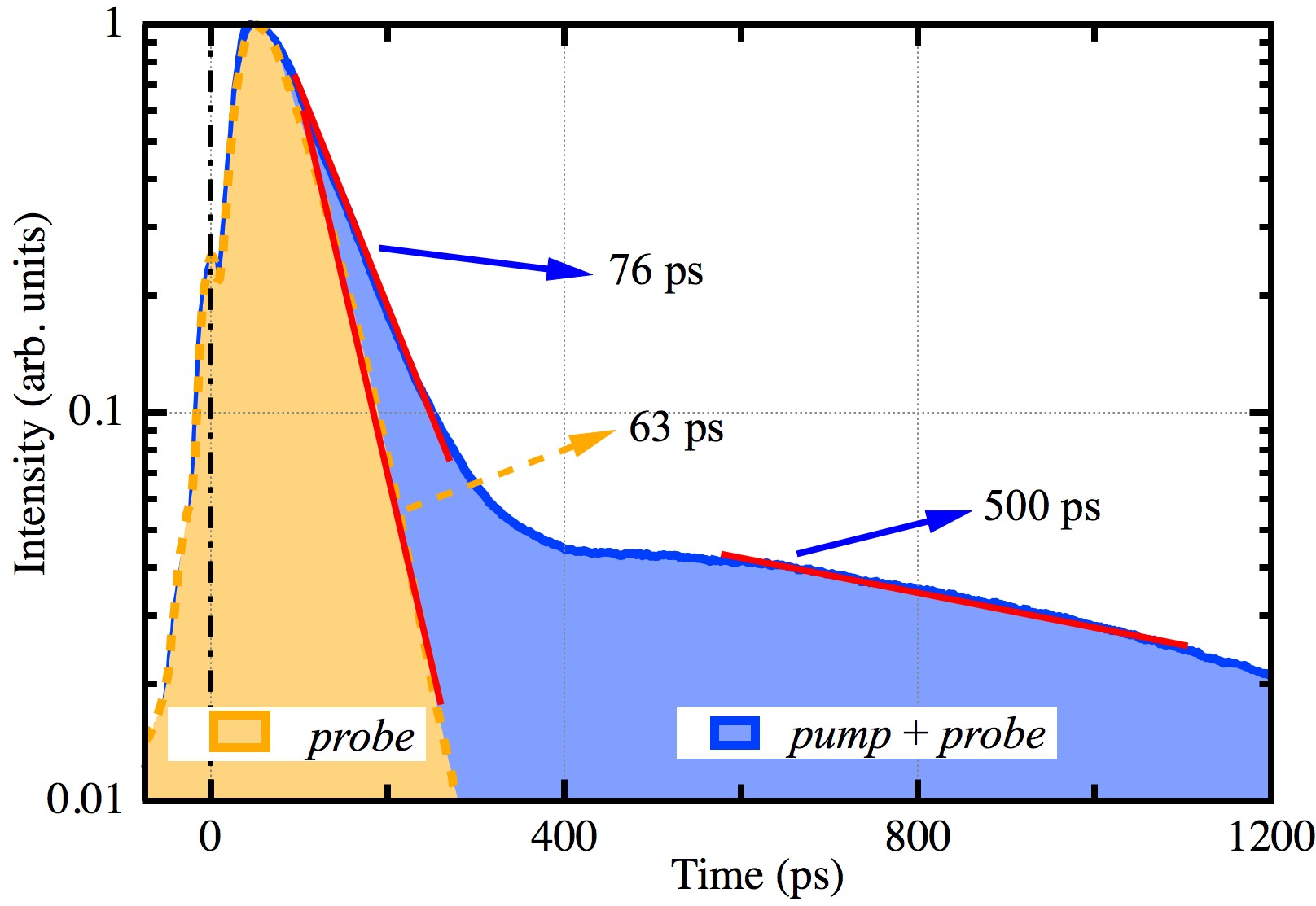}
\end{center}
\caption{(Color online) Dynamics of the integrated polariton PL under the two excitation conditions: \emph{pump}+\emph{probe} (full line) and \emph{probe}-only (dashed line). Full, straight lines are included in the different exponential decays as guides to the eye. PL is plotted in a logarithmic axis.}
\label{fig:int_prof}
\end{figure}

The dynamics of the spatially integrated emission build up is depicted in Fig.~\ref{fig:int_prof}. A similar behavior is obtained for \emph{pump}+\emph{probe} (blue line) and \emph{probe}-only (orange line) excitation conditions. The PL reaches its maximum emission $\sim$50 ps after the \emph{probe} is gone. It is on the decay dynamics that differences between the two excitation conditions appear. Under \emph{probe}-only excitation we observe a mono-exponential decay of the PL, with a characteristic decay time of 63 ps. The decay is markedly different in the presence of the out-of-resonance \emph{pump}. In this case, the PL decay is bi-exponential, with a fast decay time of 76 ps, and a long-lived polariton population lasting more than 1 ns, evidenced by the slow, $\sim500$ ps, decay time.

\section{Theoretical description}

To model our experimental results under the OPO configuration, described in Section~\ref{subsec:p+pb}, we make use of the 2D coupled Gross-Pitaevskii equations for photons $\psi(x,y)$, Eq.~\ref{eq:GPE1}, and excitons $\varphi(x,y)$, Eq.~\ref{eq:GPE2}:

\begin{equation}
\begin{gathered}i\hbar\frac{{\partial\psi}}{{\partial t}}=-\frac{{\hbar^{2}}}{{2m_{ph}}}\Delta\psi+\frac{{\hbar\Omega_{R}}}{2}\varphi\\
+U\psi-\frac{{i\hbar}}{{2\tau_{ph}}}\psi+P+P_{X}+f
\end{gathered}
\label{eq:GPE1}
\end{equation}

\begin{equation}
\begin{gathered}i\hbar\frac{{\partial\varphi}}{{\partial t}}=-\frac{{\hbar^{2}}}{{2m_{X}}}\Delta\varphi+\frac{{\hbar\Omega_{R}}}{2}\psi+\alpha_{1}\left|{\varphi}\right|^{2}\varphi\\
+U\varphi
\end{gathered}
\label{eq:GPE2}
\end{equation}

Here, $m_{ph}=4\times10^{-5}m_{0}$ is the photon mass, $m_{X}=0.6m_{0}$ is the exciton mass ($m_{0}$ is the free electron mass),
$\hbar\Omega_{R}=9$ meV is the Rabi splitting, $\alpha_{1}=6 E_{b} a_{B}^2$ is the triplet interaction constant~\cite{Vladimirova:2010} ($E_{b}=10$ meV is the exciton binding energy and $a_{B}=10$ nm is the exciton Bohr radius).

The confinement potential of the pillar, acting on the photonic and excitonic parts, is described by $U$. $\tau_{ph}=1$ ps is the photon lifetime (the exciton decay is neglected), $P$ is the quasi-resonant pumping term, exciting the system at a given frequency and in-plane momentum (same values as in the experiments), blue-detuned with respect to the polariton branch, and $f$ is the noise, which serves to account for the effects of spontaneous scattering. The \emph{cw} pumping provides an average of 10 particles in a unit cell of $h=0.25$ $\mu$m in the steady state, while the spontaneous scattering creates 0.01 particles. To describe the non-resonant \emph{probe}, we use a pulsed pumping term $P_{X}$, tuned at the exciton resonance, with the same duration and wavevector as in the experiments. No disorder potential was taken into account, because its effects were not observed in the experiments.

\begin{figure}[!htb]
\setlength{\abovecaptionskip}{-5pt}
\setlength{\belowcaptionskip}{-2pt}
\begin{center}
\includegraphics[trim=0.0cm 0.5cm 0.2cm 0.0cm, clip=true,width=1.0\linewidth,angle=0]{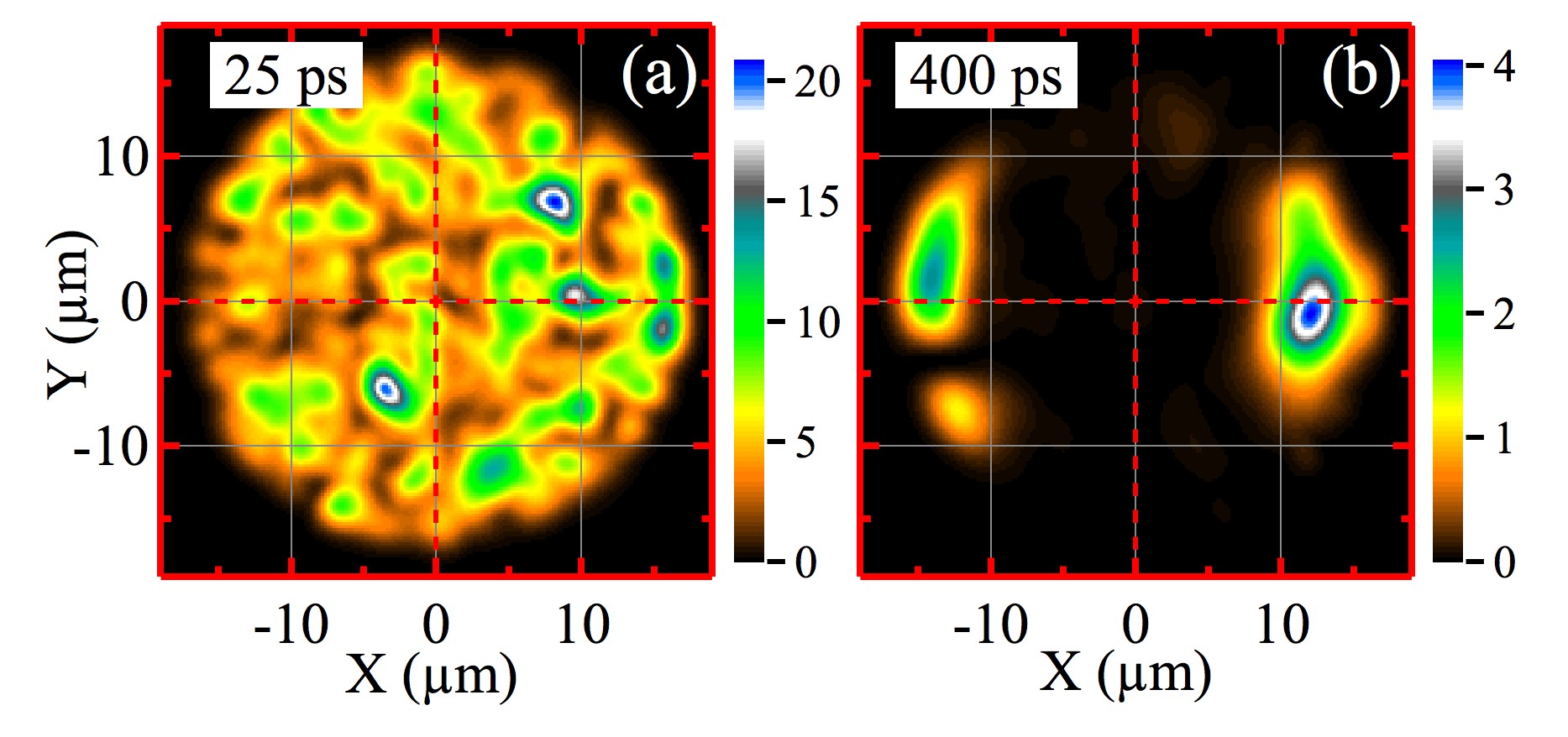}
\end{center}
\caption{(Color online) Real-space polariton emission dynamics in the pillar under \emph{pump} and \emph{probe} beam excitation. Panel (a)/(b) correspond to 25/400 ps after the \emph{probe} excitation. The PL is coded in a linear, false color scale.}
\label{fig:theory}
\end{figure}

The results of the theoretical simulations are presented in Fig.~\ref{fig:theory}. We plot the spatial distribution of the PL from the MC for the \emph{signal} state, by applying a filter in $\mathbf{k}$-space, blocking the emission for $\mathbf{k} > 1.5$~$\mu$m$^{-1}$ as in the experiments. This spatial distribution is plotted at two different instants, in order to demonstrate the agreement between experiment and theory. Figure~\ref{fig:theory}(a)/(b) (25/400 ps delay) is to be compared with Fig.~\ref{fig:realCWpulsed}(c)/(g).

First of all, the main conclusion is that both in experiment and theory the creation of excitons by the \emph{probe} laser leads to a blue-shift of the polariton dispersion, which moves it into resonance with the \emph{pump} laser. The density of the \emph{pump} state becomes then sufficient to start OPO scattering into the \emph{signal} and \emph{idler} states.

Because of the hysteresis of the polariton system in the bistable regime, the \emph{pump} density remains high even when the probe excitons have decayed through relaxation and emission from strongly photonic states. The OPO regime persists for more than 1 nanosecond, much longer than the characteristic decay times in the system. The presence of the confinement potential of the pillar prevents the escape of the \emph{signal} polaritons under the influence of the potential of the \emph{pump}, contributing to the persistence of the OPO. The decay of the OPO is due to the overall imbalance between pumping and losses. The small difference between pumping and losses becomes more and more important because the decrease of the total polariton density leads to a redshift of the pumped mode, bringing it out of OPO resonance, decreasing the efficiency of pumping even further. This is evidenced in the experiments by the fact that the time evolution of the intensity becomes a concave function of time (the effective lifetime becomes progressively shorter at longer times). In the simulations, increasing the pumping allows to overcome this decay and to maintain a permanently stable OPO regime. However, one should note that in the bistable configurations, the presence of noise (spontaneous scattering) leads to spontaneous transitions between the two stable solutions at long time scales. This noise also contributes to drive the system towards the low-intensity state and to switch-off the OPO.

More conclusions can be drawn from the spatial shape of the signal at different moments of time, which results from the interplay of several effects. At initial moments of time, the density becomes sufficient to surpass the OPO threshold everywhere in the pillar, with a large spreading in the reciprocal space caused by different phase matching conditions (different signal wave vectors and energies) in different points because of the non-homogeneous pump profile. However, the \emph{signal} exhibits a homogeneous emission over the whole surface of the pillar, the rapid spreading of polaritons being caused by the strong interactions between all particles: those created by the \emph{pump} (once the bistability is passed) and those injected with the \emph{probe}. This is well reproduced by the theoretical simulations [Fig. ~\ref{fig:theory}(a)]. At later times [Fig. ~\ref{fig:theory}(b)], as the total density drops down and the spatial redistribution stabilizes, the signal concentrates around opposite sides of the pillar along the X axis. This is due to the spatial shape of the \emph{pump} laser, which creates the conditions favorable for the OPO only in these points, located at the minima between the \emph{pump} spot and the pillar boundary: the \emph{pump} density at the center of the pillar is too high to maintain resonant OPO, because the signal polaritons are pushed away from the center and there their density becomes insufficient to maintain stable OPO. The overall good agreement between theory and experiment supports our interpretation of the experimental observations.

\section{Conclusions}
\label{sec:conclusions}

We have presented new experimental conditions to obtain a long-lived polariton condensate in a pillar MC. It involves two excitation beams impinging at the center of the pillar with the same wavevector: a \emph{cw} \emph{pump}, slightly blue-detuned from the inflection point of the LPB, and a pulsed \emph{probe}, resonantly creating excitons. The polariton population created with the arrival of the \emph{probe} induces a blue-shift of the LPB, which enters into resonance with the \emph{pump} beam, triggering the OPO-process. The \emph{cw} \emph{pump} keeps on feeding the OPO after the pulsed \emph{probe} has disappeared, because of the hysteresis of the polariton bistability. As a result of the combined effect of both beams, the OPO signal emission lasts for more than 1 ns, much longer than any of the characteristic times of the MC. The polariton condensate dynamics observed when using just the \emph{probe} beam is remarkably similar, but much shorter lived, to that obtained for the two beam excitation. The exciton population created by the \emph{probe} beam efficiently relaxes to the LPB and from there follows the OPO dynamics. The characteristic decay time is one order of magnitude shorter than that obtained under the two beam excitation conditions.
	

\section{Acknowledgements}
\label{sec:acknow}
	
C.A. acknowledges financial support from Spanish FPU scholarship. P.S. acknowledges Greek GSRT program ``ARISTEIA'' (1978) and EU ERC ``Polaflow'' for financial support. The work was partially supported by the Spanish MEC MAT2011-22997 and EU ITN INDEX (289968) projects.


%


\begin{thebibliography}{30}
\bibitem [{det()}]{detunning}%
  \BibitemOpen
  \href@noop {} {\bibinfo  {journal} {We define the exciton--photon detuning as $\delta=E_C-E_X$, where $E_C$ and $E_X$ are the cavity mode and exciton energy at $\mathbf{k}=0$, respectively}\ }\BibitemShut {NoStop}%
\end{thebibliography}
\end{document}